\pdfoutput=1

\documentclass[12pt]{article}
\usepackage{graphicx}
\usepackage{epsfig}
\usepackage[T1]{fontenc}
\usepackage{floatrow}
\newcommand\pubnumber{NuPhys2015-Tamborra}
\newcommand\pubdate{\today}

\def\ksu{GRAPPA Institute, University of Amsterdam\\
Science Park 904, 1098XH Amsterdam, The Netherlands}
\def\support{\footnote{Current institution: Niels Bohr Institute, University of Copenhagen, Blegdamsvej 17, 2100 Copenhagen, Denmark.}} 

\def\Title#1{\begin{center} {\Large #1 } \end{center}}
\def\Author#1{\begin{center}{ \sc #1} \end{center}}
\def\Address#1{\begin{center}{ \it #1} \end{center}}

\newcommand\pubblock{\rightline{\begin{tabular}{l} \pubnumber\\
         \pubdate  \end{tabular}}}
\newenvironment{Abstract}{\begin{quotation}  }{\end{quotation}}
\newenvironment{Presented}{\begin{quotation} \begin{center} 
             PRESENTED AT\end{center}\bigskip 
      \begin{center}\begin{large}}{\end{large}\end{center} \end{quotation}}
\def\Acknowledgements{\bigskip  \bigskip \begin{center} \begin{large}
             \bf ACKNOWLEDGEMENTS \end{large}\end{center}}

\def\beq{\begin{equation}}
\def\eeq#1{\label{#1}\end{equation}}
\def\eeqn{\end{equation}}

\def\beqa{\begin{eqnarray}}
\def\eeqa#1{\label{#1}\end{eqnarray}}
\def\eeqan{\end{eqnarray}}

\let\bar=\overbar

\def\Dslash{\not{\hbox{\kern-4pt $D$}}}
\def\dslash{\not{\hbox{\kern-2pt $\del$}}}

\def\msb{{\bar{\ssstyle M \kern -1pt S}}}

\begin{document}
\begin{titlepage}
\pubblock

\vfill
\Title{Supernova Neutrinos: Theory}
\vfill
\Author{Irene Tamborra\support}
\Address{\ksu}
\vfill
\begin{Abstract}
Neutrinos play a key role in core-collapse supernova explosions. Carrying information from deep inside the stellar core, neutrinos are direct probes of the supernova mechanism. Intriguing recent developments on the role of neutrinos in supernovae are reviewed, as well as our current understanding of the flavor conversions in the stellar envelope, and the detection perspectives of the next burst.\end{Abstract}
\vfill
\begin{Presented}
NuPhys 2015, Prospects in Neutrino Physics\\
Barbican Center, London, UK, December 16--18, 2015
\end{Presented}
\vfill
\end{titlepage}
\def\thefootnote{\fnsymbol{footnote}}
\setcounter{footnote}{0}

\section{Introduction}
The only core-collapse supernova (SN) detected in neutrinos has been the SN 1987A; although only few neutrinos events were collected in that occasion, this burst has been the first proof of the stellar collapse and revealed to be extremely useful to probe  the neutrino properties as well as non-standard physics scenarios. 

Since the SN 1987A, substantial progress has been
made as for   our understanding of the physics leading to the core collapse~\cite{Janka:2012wk,Janka:2016fox}, the role of neutrinos in the star, and  the flavor oscillations  in neutrino-dense media~\cite{Mirizzi:2015eza,Chakraborty:2016yeg}. 
Several large scale detectors are in place (or will be soon) to detect the next galactic burst~\cite{Mirizzi:2015eza,Scholberg:2012id}.  Nevertheless, we are still far from fully grasping the core collapse physics and the role of neutrinos in it. In this sense, the detection of the next galactic burst will provide us with a precious test of our understanding of the collapse dynamics.

Besides the single SN burst, the detection of the diffuse SN neutrino  background (DSNB), the neutrino flux emitted from all SNe exploding somewhere in the Universe, is approaching. The DSNB will allow us to learn about the stellar population, other than provide with an independent test of the SN rate~\cite{Mirizzi:2015eza,Beacom:2010kk,Lunardini:2010ab}. 

In what follows, we will outline some of the open issues in SN  and neutrino astrophysics. The detection perspectives of the next burst will be briefly discussed.    

\section{Core-collapse supernova physics}
A core-collapse SN explosion originates from the death of a massive star ($M> 8 M_\odot$)~\cite{Janka:2012wk}. The SN iron core is surrounded by shells of lighter elements; once the Chandrasekhar limit  is reached, the core collapses and the explosion is triggered. The $99\%$ of the explosion energy is released in neutrinos with average energies of $\mathcal{O}(10)$~MeV for about 10~s. 

\begin{figure}[t]
\centering
\epsfig{figure=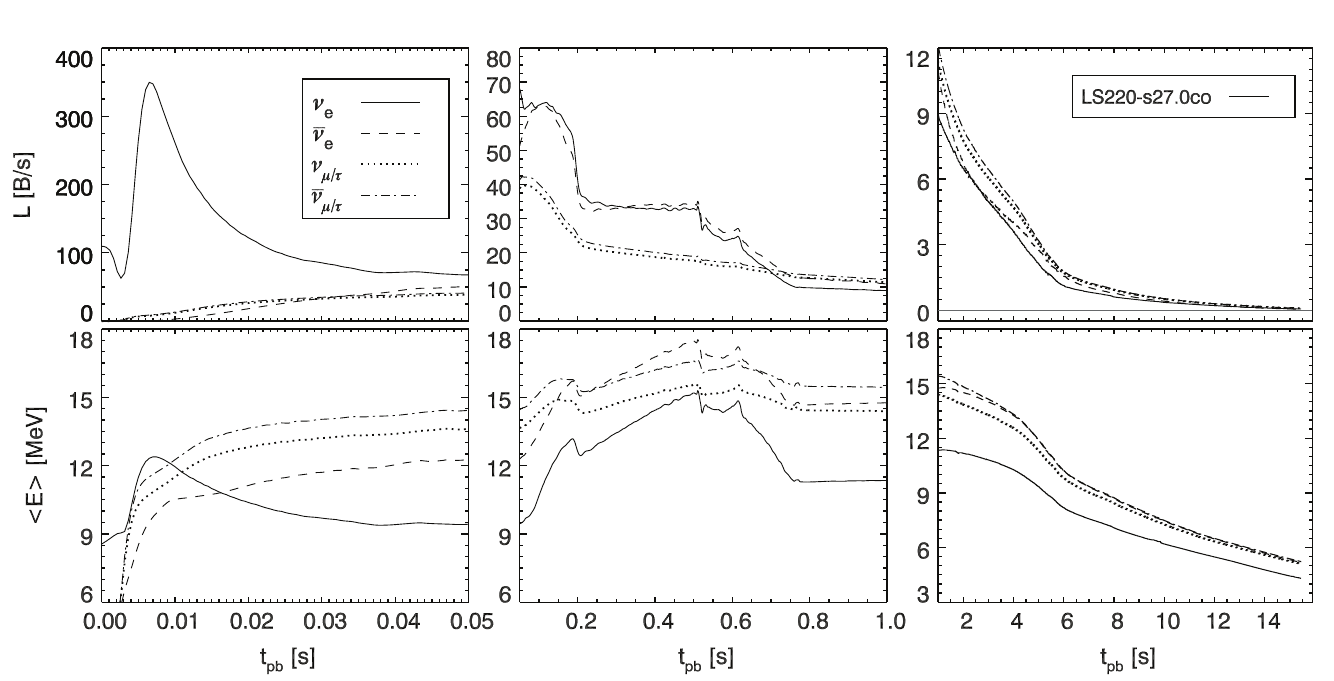,width =1.05\columnwidth}
\caption{Neutrino luminosities (on the top) and mean energies (on the bottom) for the three flavors [continue line stands for $\nu_e$'s, dashed line for $\bar{\nu}_e$'s, dotted (dashed-dotted) line for $\nu_{\mu,\tau}$ ($\bar{\nu}_{\mu,\tau}$)] as a function of the post-bounce time for a 27 $M_\odot$ SN progenitor. The panels on the left refer to  the neutronization phase, the middle panels to the accretion phase, and the panels on the right to the cooling phase. Figure adapted from from Ref.~\cite{Mirizzi:2015eza}.
\label{fig1}}
\end{figure}

The SN neutrino signal can be divided in three main phases, as shown in Fig.~\ref{fig1}: The {\emph{neutronization burst}} marked by a large peak in the $\nu_e$ luminosity (generated because of the rapid electron capture by nuclei and  free protons, as the shock wave crosses the iron core dissociating its nuclei); the {\emph{accretion phase}} where the differences among the fluxes of different flavors are still large especially between the electron and non-electron flavors; the {\emph{cooling phase}} where the neutrino emission properties among the different flavors became very similar and the luminosity progressively decreases.

Core-collapse hydrodynamic simulations have recently reached the 3D front, unveiling new and unexpected features~\cite{Janka:2016fox}. The first successful explosions have been obtained in 3D~\cite{Melson:2015spa,Melson:2015tia}. For example, in one of these cases, the explosion was triggered through a $10\%$ reduction of the neutrino opacity, obtained by adding a strange quark contribution to the nucleon spin~\cite{Melson:2015spa}. Such correction might be optimistic with respect to the current bounds; however this explosion, obtained by slightly perturbing the initial conditions, suggests we might be close to fully reproduce  the core-collapse conditions. 

\begin{figure}[t]
\centering
\epsfig{figure=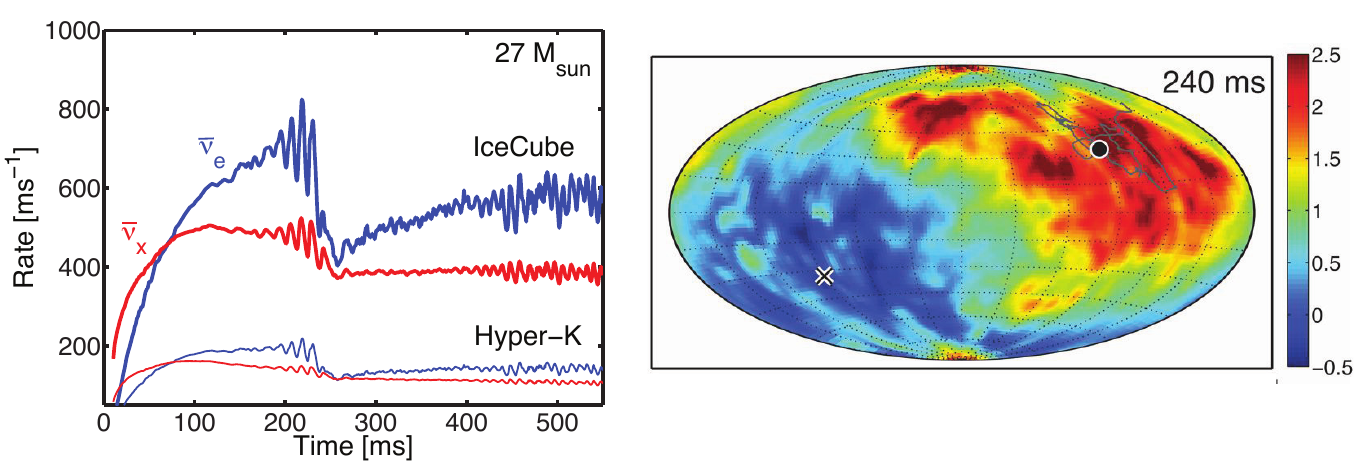,width =1.05\columnwidth}
\caption{Left: Detection rate in IceCube and Hyper-Kamiokande as a function of the post-bounce time for a 27 $M_\odot$ SN progenitor
for an observer at 10~kpc located along a direction where the neutrino signal shows strong SASI modulations. Right: Neutrino
lepton number flux ($\nu_e - \bar{\nu}_e$) normalized by its average value for a $11.2\ M_\odot$ progenitor
at $t_{\mathrm p.b.}=240$~ms. 
Figure adapted from from Refs.~\cite{Tamborra:2014hga,Tamborra:2014aua}.
\label{fig2}}
\end{figure}

The SN explosion is expected to occur according to the delayed neutrino explosion mechanism~\cite{Bethe:1984ux}: The shock wave looses all its energy while propagating through the iron core and dissociating iron nuclei. Neutrinos provide fresh energy to revive the stalled shock. During the shock revival phase hydrodynamical instabilities occur, such as convective overturns and the standing accretion shock instability (SASI) that contribute to enhance the efficiency of the energy transfer between the neutrinos and the shock wave.  Recent 3D SN simulations suggest that the neutrino signal carries imprints of such instabilities~\cite{Tamborra:2014hga,Tamborra:2013laa}. For example, SASI episodes are expected to occur in the heavy mass SN progenitors and will be clearly detectable with neutrino telescopes such as IceCube and Hyper-Kamiokande as shown in the left panel of Fig.~\ref{fig2} for a 27 $M_\odot$ SN progenitor. Among the well known hydrodynamical  instabilities, yet another instability has been recently discovered: The lepton emission self-sustained asymmetry (LESA)~\cite{Janka:2016fox,Tamborra:2014aua}; LESA is the first instability driven by neutrinos and it consists of an asymmetric emission of the $\nu_e$ number flux with respect to the $\bar{\nu}_e$ one, see the right panel of Fig.~\ref{fig2}. LESA is characterized by a large scale dipolar character  and, once it develops, its axis remains stable despite SASI or convective motions. LESA is  responsible for a strong directional dependence of the neutrino fluxes that could affect the SN nucleosynthesis, oscillations and neutron star kicks. 

\section{Neutrino flavor oscillation physics}
The flavor evolution of neutrinos is described by matrices of densities 
for each energy  mode $E$, the 
diagonal entries being the occupation numbers. The evolution of  $\rho_E$ is described by the Liouville equations
\begin{equation}\label{eq:eom1}
\mathit{i}\partial_r\rho_E=[{\sf H}_{E},\rho_{E}]
\quad\hbox{and}\quad
\mathit{i}\partial_r\bar\rho_E=[\bar{\sf H}_{E},\bar\rho_{E}]\,,
\end{equation}
where an overbar has been adopted to mark the $\bar{\nu}$ quantities. Each matrix  $\rho_E$ is a  $3{\times}3$ matrix in the flavor space.
The Hamiltonian matrix contains vacuum, matter, and neutrino--neutrino terms:
${\sf H}_{E}= {\sf H}^{\rm vac}_{E}+{\sf H}^{\rm m}_{E}+{\sf H}^{\nu\nu}_{E,\vartheta}$; in particular,  ${\sf H}^{\nu\nu}$ depends on the 
angle between the momenta of the colliding neutrinos for each energy mode $E$.

The resonant flavor conversions due to the interactions of the neutrinos with the matter background in the stellar envelope (MSW effect)
is a  well understood phenomenon~\cite{Mikheev:1985,Dighe:1999bi}. The first MSW resonance is due
to the atmospheric neutrino mass difference and it occurs at about $10^3$~km from the neutrino sphere in the neutrino (antineutrino) channel for  normal (inverted) mass ordering. The second MSW conversion takes place at larger radii and it is due to the smaller solar neutrino mass difference.
 
 Supernovae are neutrino-rich environments and therefore $\nu$--$\nu$ interactions cannot be neglected~\cite{Mirizzi:2015eza,Chakraborty:2016yeg,Duan:2010bg}. 
 Neutrino self-interactions are non-linear effects, inducing exponentially growing instabilities
 in the flavor space. The angle between the momenta of the colliding neutrinos  
is extremely important in defining the final distribution of the neutrino fluxes. 

The modelling of such interactions has been first developed 
within the  so-called {\emph{neutrino bulb model}}~\cite{Duan:2006an}:  Neutrinos
of all flavors are emitted from each point of the neutrino-sphere in the forward solid angle uniformly and isotropically. The most famous signature
of $\nu$--$\nu$ interactions is the {\emph{spectral split}}: A complete swap of the neutrino energy spectra
for certain energy ranges and according to the neutrino mass hierarchy, as shown in Fig.~\ref{fig2}. 
\begin{figure}[t]
\centering
\vspace*{-4mm}
\hspace*{-6mm}
\epsfig{figure=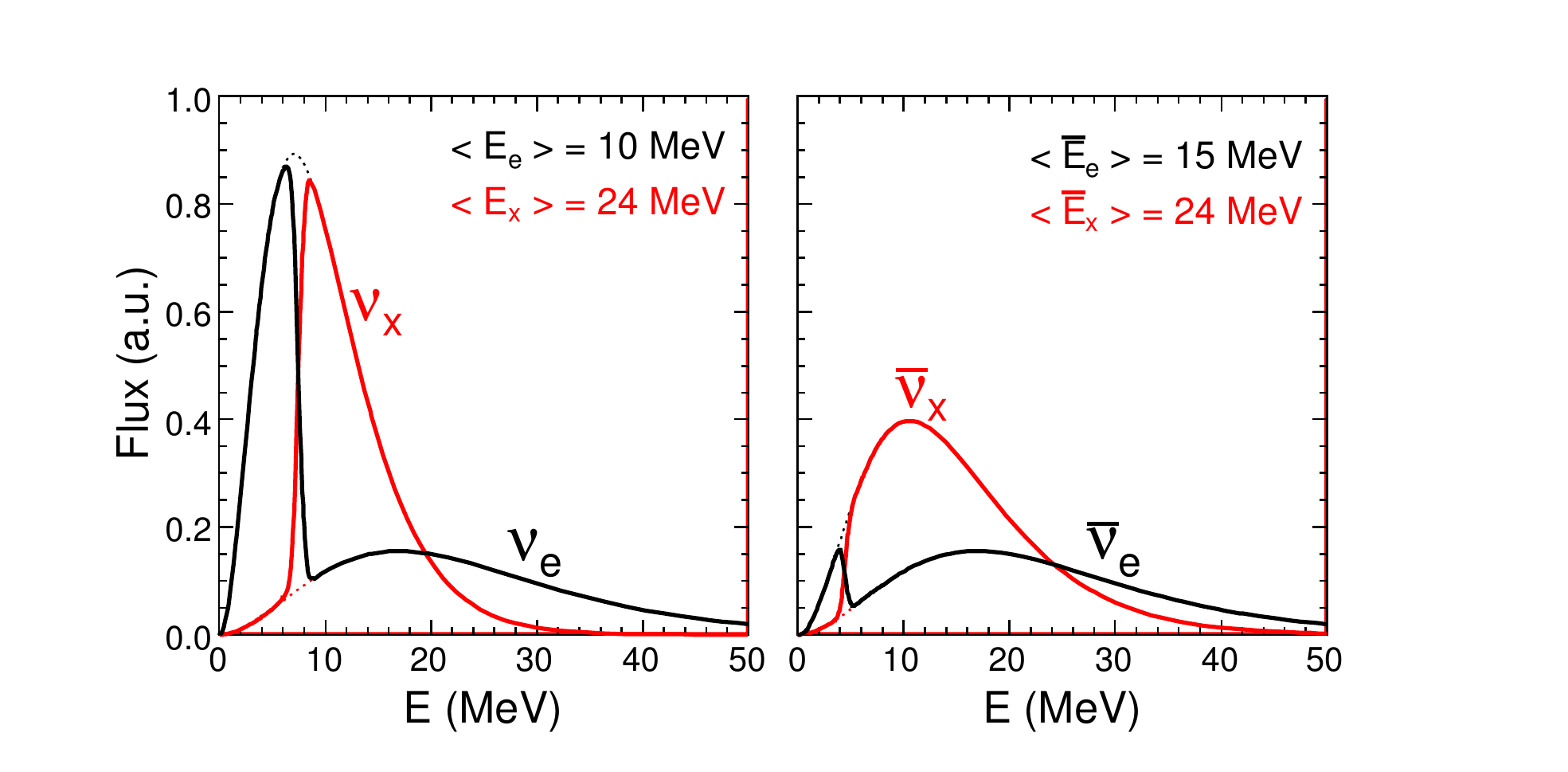,width =0.8\columnwidth}
\caption{Neutrino (on the left) and antineutrino fluxes (on the right) as a function of the energy after $\nu$--$\nu$ interactions assuming inverted mass ordering. The fluxes  at the neutrino sphere are shown
as dotted lines. Figure taken from Ref.~\cite{Fogli:2008pt}.
\label{fig3}}
\end{figure}

One would expect that the impact of flavor conversions in determining the final neutrino energy distributions should be of relevance during the early phase of the SN signal where the neutrino emission properties  among the different flavors are larger. However, it has been proved that the high matter density profile, typical of these post-bounce times, locks the neutrino modes inhibiting multi-angle flavor conversion effects ($N_e \ge N_\nu$)~\cite{EstebanPretel:2008ni}. Such prediction has been numerically proved  within  simplified setups and also analytically relying on the stability analysis approach.  On the other hand, during the cooling phase, where the fluxes of the neutrinos of different flavors are more similar to each other, multiple spectral splits are expected to occur~\cite{Fogli:2009rd}.

The above conclusions have been drowned by relying on the assumption that we have a stationary, spherically symmetric SN, where the neutrino fluxes evolve with radius. However, more recently, it has been pointed out as by releasing such approximations, new instabilities in the flavor space may arise. 
For example, within a simplified setup, it has been shown as breaking the axial symmetry~\cite{Raffelt:2013rqa}, the spatial and directional symmetry~\cite{Duan:2014gfa}, or by introducing temporal instabilities~\cite{Abbar:2015fwa},  flavor conversions could be induced (i.e., flavor instabilities can be determined because of the non-homogeneous or non-stationary conditions occurring within the stellar envelope). The same
should be expected by considering a neutrino angular distribution not limited to the outward direction, as well as in the presence of large 3D effects that make the system inhomogeneous, non-stationary and anisotropic~\cite{Tamborra:2013laa,Tamborra:2014aua}. Existing  investigations in this contest are
 still simplified cases of study and further work is necessary.

\section{What can we learn from the next SN burst?}
If we assume that even within a more realistic modelling of the flavor oscillations,  the matter potential mostly suppresses  $\nu$--$\nu$ multi-angle matter
effects during the accretion phase, while  neutrino self-interactions are responsible for multiple spectral splits during the cooling phase, then each of the three phases of the SN neutrino signal offers different opportunities to learn about the stellar collapse or neutrino properties.

The neutronization burst signal is independent of the progenitor mass and the nuclear equation of state. It can be adopted as a standard candle to define the distance of the SN event~\cite{Kachelriess:2004ds}. Since the slope of the $\bar{\nu}_e$ and  $\bar{\nu}_x$ light-curves  is different, the observed neutrino event rate will be sensitive to the neutrino mass ordering in, e.g., Cherenkov telescopes. The same holds for the neutrino channel; if we consider the $\nu_e$ event rate as seen in e.g.,~a liquid argon detector, the absence (presence) of the peak of the neutronization burst will hint towards a normal (inverted) mass ordering~\cite{Wallace:2015xma}. 

During the accretion phase, the neutrino signal is dependent on the neutrino mass hierarchy. Moreover, it carries characteristic signatures of the SASI motions and convective overturns, clearly detectable in, e.g., Cherenkov telescopes~\cite{Tamborra:2013laa,Tamborra:2014hga} providing insights on the core-collapse physics.  

The cooling phase signal is strongly sensitive to the nuclear equation of state as well as to the SN progenitor mass. The exact composition in neutrinos of different flavors  is responsible for determining the nucleosynthesis outcome in the neutrino driven wind~\cite{Duan:2010af,Pllumbi:2014saa}.  Although recent work suggests that, even by taking into account  the existence of an extra light sterile family, it is difficult to create a n-rich environment in the SN neutrino driven wind and to  activate the r-process~\cite{Pllumbi:2014saa}, the role of oscillations in the production of heavy elements remains to be clarified.

\section{Diffuse Supernova Neutrino Background}
On average a SN explodes every second in the Universe and we could detect the cumulative neutrino  flux, the DSNB~\cite{Mirizzi:2015eza,Beacom:2010kk,Lunardini:2010ab}. The DSNB should be clearly detectable in the region around 20--30~MeV, where it is   above the reactor and atmospheric backgrounds. 

A class of SNe that may considerably enhance the expected DSNB  is the one of the failed progenitors~\cite{Lunardini:2009ya}. A failed SN occurs when a SN  collapses in a black hole and the neutrino signal abruptly ends after a few hundreds of ms. Preliminary work shows that the fraction of failed progenitors might be higher than thought in the past (reaching up to the 30$\%$ of the total SN population) and even light SN progenitors might generate black-hole forming cases~\cite{Ertl:2015rga}. If such predictions on black-hole forming progenitors should be confirmed, then we should expect an enhanced DSNB rate of events. 

The detection of the DSNB will be instrumental to constrain the stellar population, it will provide us with an independent test of the SN rate, and will help us to constrain the fraction of failed versus core-collapse cases. At the same time, the DSNB detection could help us to constrain the neutrino emission properties as well as  exotic physics scenarios. 

\begin{figure}[t]
\centering
\epsfig{figure=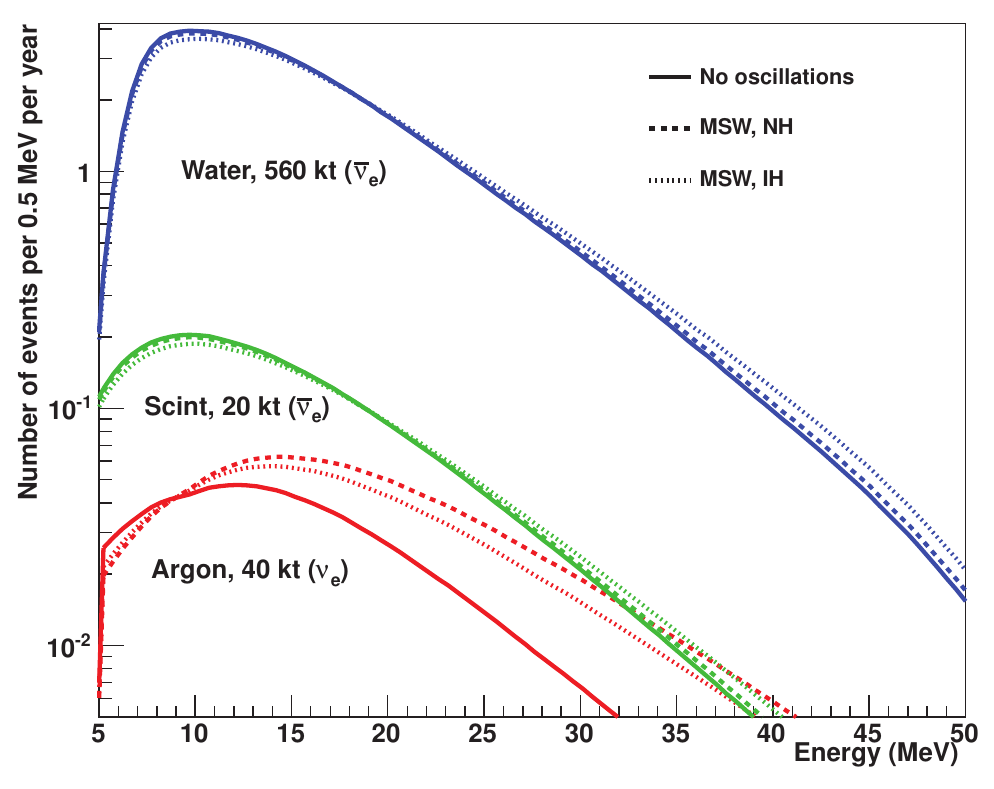,width =.6\columnwidth}
\caption{Event rates for water Cherenkov, liquid Argon and scintillator detectors as a function of the detected energy, oscillation effects included. Figure taken from from Ref.~\cite{Mirizzi:2015eza}.
\label{fig4}}
\end{figure}

The possibility of detecting the DSNB will be improved in the next future with the planned JUNO scintillator detector as well as with the approval of the Gd project for the Super-K detector~\cite{Mirizzi:2015eza}. See Figure~\ref{fig4} for an estimation of the number of events in future generation detectors.

\section{Summary}
Neutrinos play a fundamental role in the physics of a core-collapse supernova. The first supernova hydrodynamic simulations in 3D revealed unexpected and fascinating phenomena and proved as the detection of the supernova neutrino signal will be instrumental to test the explosion mechanism. 

Core-collapse supernovae are neutrino-dense environments and therefore, $\nu$--$\nu$ interactions cannot be neglected. A careful modelling of the SN environment for studying the oscillation phenomenology is compulsory  and it still incomplete at the moment, despite the intense theoretical activity in this direction. Neutrino self-interactions are non-linear effects and it has been shown as, by releasing some of the currently adopted symmetry assumptions, instabilities in the flavor space could be induced. 

Each phase of the core collapse neutrino signal could offer different opportunities to learn about the supernova physics and the synthesis of the new elements. The detection of the DSNB is expected to happen within the next decade and will offer us with a chance to constrain the stellar population as well as to independently test  the supernova rate. 

\Acknowledgements
The author warmly thanks the NuPhys 2015 organizers for their kind invitation.

\end{document}